# Non-invasive estimation of the metabolic heat production of breast tumors using digital infrared imaging

**Francisco Javier González**

*Coordinación para la Innovación y la Aplicación de la Ciencia y la Tecnología,*
*Universidad Autónoma de San Luis Potosí,*
*Sierra Leona 550, Lomas 2da. Sección, 78210,*
*San Luís Potosí, SLP, México*
*javier.gonzalez@uaslp.mx*

ABSTRACT: In this work the metabolic heat generated by breast tumors was estimated indirectly and noninvasively from digital infrared images and numerically simulating a simplified breast model and a cancerous tumor, this parameter can be of clinical importance since it has been related to the doubling volume's time and malignancy for that particular tumor. The results indicate that digital infrared imaging has the potential to estimate in a non-invasive way the malignancy of a tumor by calculating its metabolic heat generation from bioheat thermal transfer models.

# 1. Introduction

Breast cancer is the most common form of cancer among women and the second most common cancer in the World (Nover *et al.* 2009). Survival is directly related to stage at diagnosis, therefore the current approach to this disease involves early detection and treatment (Nover *et al.* 2009).

Tumors generally have an increase in blood supply and angiogenesis, as well as an increased metabolic rate which generates an increase in temperature that can be detected using digital infrared imaging (Arora *et al.* 2008).

Digital infrared imaging has been used in medical diagnostics since the 1960s and in 1982 it was approved by the US Food and Drug Administration (FDA) as an adjunctive tool for the diagnosis of breast cancer (Arora *et al.* 2008), since then the sensitivity of infrared imaging technology has increased substantially making it now a better tool for breast cancer diagnosis.

The doubling time of a tumor's volume can be used to predict how the cancer will evolve and what the expectancy of survival would be in the absence of therapy (Gautherie 1980), this doubling time is generally not measured since the only way of evaluating it is by using X-ray pictures during the period of natural evolution of the disease which is generally not feasible (Gautherie 1980).

This work is based on the possibility of finding out by computer simulations the breast's surface skin temperature due to the metabolic heat production of a tumor (González 2007).

In this work the metabolic heat generated by a breast tumor is estimated using digital infrared imaging and numerically simulating a simplified breast model and a cancerous tumor (González 2007), it is worth noting that this metabolic heat generated by the tumor has been linked to the tumor's volume doubling time which can potentially give an indication of its malignancy.

# 2. Method

## 2.1 Metabolic heat production of breast tumors

In a previous work it was reported that the metabolic heat production of a breast tumor remained approximately constant during the phase of exponential growth in spite of the histological and circulatory changes and that this metabolic heat production is related to the doubling time by a nearly hyperbolic law whereby the faster the tumor grows the more heat it generates (Gautherie 1980).

It is worth noting that the hyperbolic law shown in Fig. 1 does not take into account important breast tumor parameters such as volume and cancer stage for example, which influence the growth speed of the tumor, however this simple equation can give a rough estimate of the malignancy of a tumor given its metabolic heat production.

Analyzing the relation for the metabolic heat production of a tumor and its doubling volume time previously reported (Gautherie 1980), the following equation (Eq. 1) can be approximated:

$$y = \frac{3000}{x}, \qquad [1]$$

where $y$ is the metabolic heat production of the tumor in mW/cm$^3$ and $x$ is the doubling time in days (Figure 1).

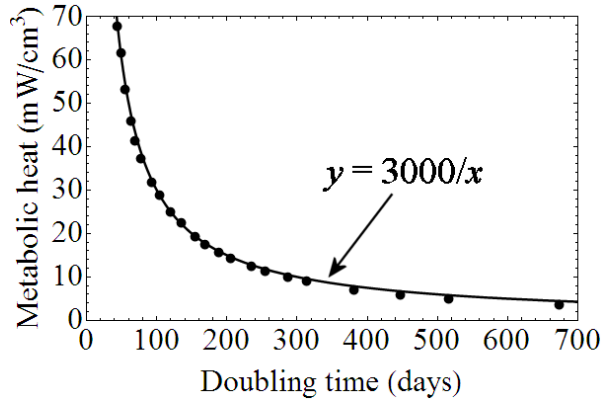

**Figure 1.** *The dots represent the relation reported (Gautherie 1980) for the metabolic heat production of a breast tumor and its volume doubling time, this relation can be approximated by the equation y = 3000/x where y is the metabolic heat production of the tumor in mW/cm$^3$ and x is the doubling time in days.*

*2.2 Patients*

Twenty female subjects with ages ranging from 35 to 81 years old (mean=54 years, SD=12 years) with confirmed diagnosis of an infiltrating ductal carcinoma and scheduled for surgery, either mastectomy or lumpectomy at the "Hospital Dr. Ignacio Morones Prieto" in San Luis Potosi, Mexico, participated in this study. Informed consent was obtained from all participants and the study was approved by the local ethics committee.

*2.3 Clinical Evaluation*

All the participants in the study were subjected to mammography and biopsy and the diagnosis in all participants was of an infiltrating ductal carcinoma with tumor sizes ranging from 1 to 10 cm (mean=3.1 cm, SD=2.1 cm) and tumor depths ranging from 0.5 to 2.6 cm (mean=1.6 cm, SD=0.6 cm).

Infrared imaging was performed on all patients one day before their surgery and all the patients had their biopsies performed at least two weeks before surgery. The infrared camera used was a FLIR T400 (FLIR Systems, Wilsonville, OR) which has a 320 × 240 Focal Plane Array of uncooled microbolometers with a spectral range of 7.5 to 13 µm and a thermal sensitivity of 50 mK at 30 °C.

The participants did not engage in physical activities, smoked, drank alcohol or applied lotion to the breast within 6 hours before the infrared images were taken, they were also asked to sit disrobed for 15 minutes, before the images were taken, in a temperature controlled room maintained between 23 and 25 °C in order to stabilize their body temperature.

The infrared images were taken with the participants in the standing position, with their hands holding the back of their head and 1.5 m away from the infrared camera, five infrared images were taken for each participant, one frontal, two oblique views (left and right lateral) and a frontal close-up of each breast.

The images were viewed and analyzed using the FLIR QuickReport software (FLIR Systems, Wilsonville, OR), the interpretation of the images was done using a thermal score derived from the Ville Marie Infrared grading scale (Keyserlingk *et al.* 1998). This thermal score takes into account the two most significant infrared signs which are: (a) the difference in surface temperature at the lesion site from that at the mirror image site on the contralateral breast (ΔT), and (b) the vascular pattern at and around the lesion site, when the contralateral breast did not reveal such a pattern (Wang *et al.* 2010).

The thermal score used was calculated by adding the amount of vascularity to the difference in surface temperature in degrees centigrade at the lesion site compared to the contralateral breast (Figure 2).

The amount of vascularity was determined using the following scale:

1: Absence of vascular patterns.
2: When symmetrical or moderate vascular patterns were found.
3: In the case of significant vascular asymmetry.
4: Extended vascular asymmetry in at least one third of the breast area.

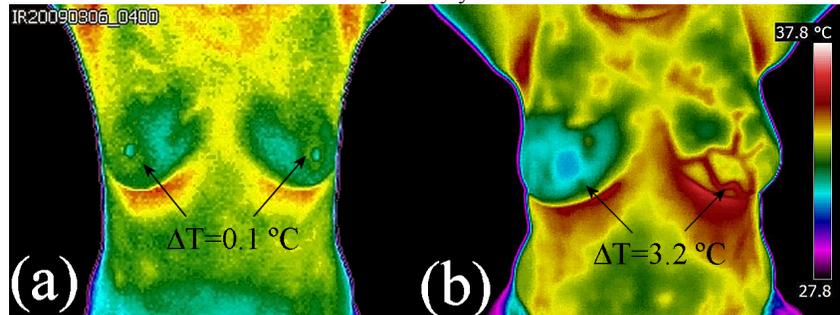

**Figure 2. (a)** *Infrared image of a female subject with healthy breasts, the calculated thermal score was 1.1, obtained by adding the amount of vascularity (1: Absence of vascular patterns) and 0.1 (difference in surface temperature, ΔT, at the lesion site compared to the contralateral breast).* **(b)** *Infrared image of a female subject with infiltrating ductal carcinoma in her left breast, mammography showed a 2 cm tumor at a depth of 1.2 cm from the skin surface, the calculated thermal score was 7.2, obtained by adding the amount of vascularity (4: Extended vascular asymmetry in at*

least one third of the breast area) and 3.2 (difference in surface temperature, ΔT, at the lesion site compared to the contralateral breast).

*2.4 Numerical simulations of a breast model and a cancerous tumor*

The mathematical model used for bioheat transfer is based on the Pennes equation (Pennes 1948) which incorporates the effects of metabolism and blood perfusion into the standard thermal diffusion equation. This equation is written in its simplified form as:

$$\rho \cdot c \frac{\partial T}{\partial t} = \nabla(k \cdot \nabla T) + \omega_b \cdot c_b \cdot \rho_b (T_a - T) + q_m , \qquad [2]$$

where $k$ is the thermal conductivity of tissue, $\rho_b$ and $c_b$ are the density and the specific heat of the blood, $\omega_b$ is the blood perfusion rate (ml/s/ml), $q_m$ is the metabolic heat generation rate (W/m$^3$), $T_a$ is the arterial blood temperature, and $T$ is the local temperature of the breast tissue. The temperature of the arterial blood is approximated to be the core temperature of the body (37 $^o$C).

The breast model used in this study was approximated using a hemisphere with a 9 cm radius to match the average patient in Gautherie's study (Gautherie 1980), also a layer 1.3 cm thick was placed beneath the hemisphere to simulate the chest wall, the tumor was considered a perfect sphere (Figure 3).

A density of 920 Kg/m$^3$ and a heat capacity of 3000 J/Kg$^o$C was used for both normal and cancerous tissue. A blood density of 1055 Kg/m$^3$ and a specific heat of 3.66×10$^3$ J/Kg·K was used in the simulations. The blood perfusion rate for normal and cancerous tissue was considered to be 0.00018 ml/s/ml and 0.009 ml/s/ml respectively (Gore and Xu 2003). An effective thermal conductivity of 0.42 W/m (Gautherie 1980) was used for both normal and cancerous tissue.

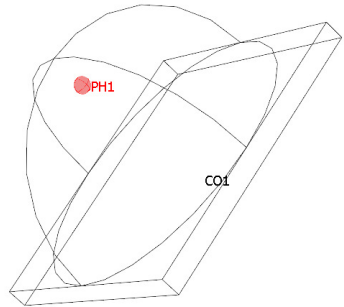

**Figure 3.** *Computational model used to replicate the typical patient in Gautherie's study. The model consists of a 9 cm radius hemisphere attached to a 1.3 cm layer that replicates the chest wall, tumors were considered to be perfect spheres.*

Finite Element Method simulations were performed using the COMSOL Multiphysics 3.5 computational package (COMSOL 2009), a convective boundary condition with a heat transfer coefficient of 5 W/m²·K was used at the skin surface to account for natural convection (Gore *et al.* 2003) and a fixed temperature boundary condition at 37 °C at the back of the chest wall was used to simulate the core temperature of the body.

An initial simulation without a tumor was performed in order to obtain the temperature distribution of the normal breast, this simulation resulted in a normal breast temperature $T_o$=32.8 °C at point *T* (Figure 4(b)). Figure 4 shows the temperature distribution obtained from finite element simulations of a breast model with a 2 cm tumor located 1.2 cm below the skin surface.

In order to obtain the metabolic heat production of a tumor a parametric simulation was performed using the size and depth of that particular tumor obtained from X-ray mammography and varying the metabolic heat term $q_m$ in Eq. 2 from a value of 0.45 mW/cm³, which is the metabolic heat production of healthy tissue (Gore *et al.* 2003), to 2000 mW/cm³ and finding the metabolic heat term that will increase the normal temperature of the breast model (32.8 °C) an amount equal to ΔT which was obtained from the clinical evaluation using digital infrared imaging.

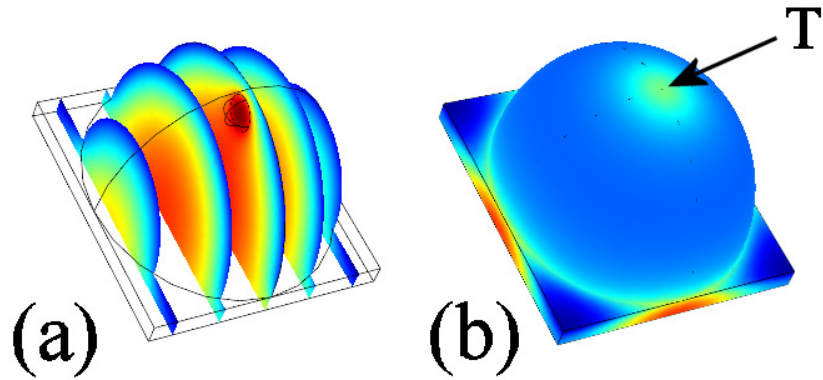

**Figure 4. (a)** *Slice graph showing the temperature distribution of the proposed breast model with a 2 cm tumor located 1.2 cm deep.* **(b)** *Temperature distribution on the surface of the proposed breast model indicating the location where the increase in temperature ΔT was measured.*

### 3. Results

The twenty participants of the present study were analyzed using digital infrared imaging using the thermal score presented in the previous section. Figure 5 shows the relation between the thermal score and the difference in surface temperature with the size of the tumor obtained by mammography. From Fig. 5(a) it can be seen that there is a significant correlation between the thermal score and the size of the tumor, however by only measuring the difference in temperature between the lesion site and the contralateral breast a stronger correlation to the size of the tumor was found (Fig. 5(b)).

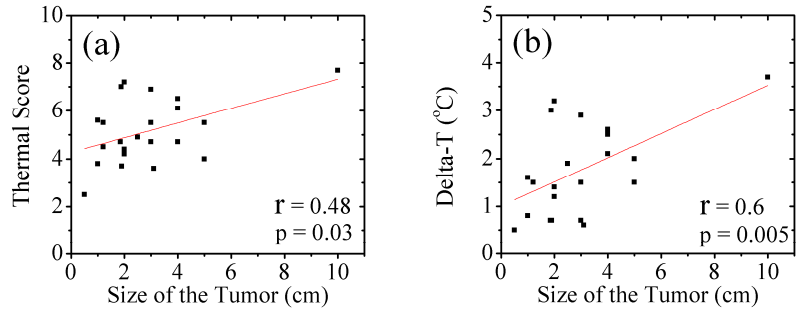

**Figure 5. (a)** *Thermal score obtained from digital infrared images of 20 patients with infiltrating ductal carcinoma as a function of the tumor size.* **(b)** *Difference in temperature of the lesion site compared to the contralateral breast as a function of tumor size.*

With the size and depth of the tumor obtained from X-ray mammography the increase in temperature at the surface of the breast can be calculated using the thermal model of a breast and a cancerous tumor presented in the previous section, the figure of merit $\Delta T$ can be obtained by also simulating the temperature of a breast without a tumor and subtracting both temperatures.

If a fixed metabolic heat generation term is used in Eq. 2 and a comparison is made between the simulations and the parameter $\Delta T$ obtained from the clinical evaluation, a significant difference between the simulated and the clinical $\Delta T$ will be encountered (Figure 6), this difference is due to the fact that each tumor has a different metabolic heat generation term.

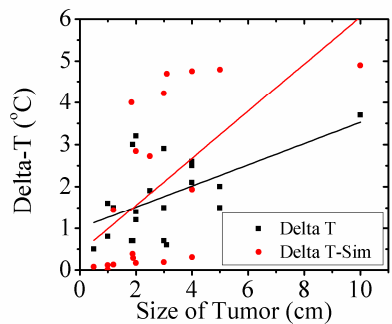

**Figure 6.** *Simulated and experimental $\Delta T$ as a function of tumor size.*

By varying the metabolic heat generation term in Eq. (2) in order to match the simulated and the experimental $\Delta T$ values the metabolic heat generated by the tumor can be estimated, using this value the doubling volume time of the tumor can also be potentially estimated.

Table I shows the tumor size and tumor depth obtained from X-ray mammography for the 20 participants of the study, also the measured ΔT values and thermal scores obtained from digital infrared imaging and the calculated metabolic heat production are presented.

| Patient # | Age (years) | Tumor Size (cm) | Tumor Depth (cm) | Measured Delta-T ($^oC$) | Thermal Score | Metabolic Heat Production ($mW/cm^3$) |
|---|---|---|---|---|---|---|
| 1 | 53 | 4 | 2.5 | 2.6 | 4.7 | 1371 |
| 2 | 58 | 2 | 2.6 | 1.2 | 4.2 | 1096 |
| 3 | 55 | 1.2 | 1 | 1.5 | 5.5 | 272 |
| 4 | 42 | 5 | 1.05 | 1.5 | 5.5 | 96 |
| 5 | 55 | 5 | 2 | 1.5 | 5.5 | 376 |
| 6 | 35 | 3 | 1.5 | 1.5 | 5.5 | 272 |
| 7 | 65 | 3 | 1.1 | 0.3 | 3.3 | 29 |
| 8 | 69 | 2.5 | 1.5 | 1.9 | 4.9 | 399 |
| 9 | 44 | 1.2 | 2.1 | 1.5 | 4.5 | 1312 |
| 10 | 63 | 1.2 | 1 | 1.2 | 4.2 | 205 |
| 11 | 55 | 5 | 2 | 1.2 | 5.2 | 295 |
| 12 | 53 | 1.88 | 2.1 | 3 | 7 | 1645 |
| 13 | 52 | 3.1 | 0.95 | 0.6 | 3.6 | 8 |
| 14 | 56 | 1.85 | 0.95 | 0.7 | 4.7 | 44 |
| 15 | 73 | 10 | 0.5 | 4.1 | 8.1 | 75 |
| 16 | 53 | 4 | 2.5 | 2.6 | 6.6 | 1272 |
| 17 | 36 | 1.8 | 2.01 | 0.4 | 4.4 | 454 |
| 18 | 49 | 1 | 1.95 | 1.6 | 5.6 | 1464 |
| 19 | 81 | 1.9 | 2.25 | 0.7 | 4.7 | 448 |
| 20 | 36 | 4 | 0.85 | 2.1 | 6.1 | 116 |

**Table I.** *Tumor parameters obtained from X-ray mammography, digital infrared imaging and finite element simulations.*

## 4. Conclusion

Breast tumors produce an increase in temperature at the surface of the breast compared to the contralateral breast that can be detected with digital infrared imaging. This increase in temperature depends not only on the size and depth of the tumor but also on its malignancy since tumors with faster volume doubling times will generate more heat than tumors with slower doubling volume times.

Even though the thermal score presented in this work and even the parameter ΔT alone can detect the presence of tumors, the detection sensitivity greatly depends on the malignancy of the tumor. In a previous work (González 2007) it was concluded by using computer simulations that state of the art infrared cameras could detect tumors as small as 0.5 cm at a depth of 2 cm, these simulations were performed using a fixed metabolic heat generation value for the tumor, the experimental results of the current work show that tumors present very wide ranges of metabolic heat generation values which is an important factor that will determine the ultimate sensitivity of digital infrared imaging.

So far the only relationship between metabolic heat generation and malignancy of a tumor is the hyperbolic law presented in Gautherie's 1980 paper (Gautherie 1980), however from the metabolic heat values presented in this paper it is clear that even though large tumors can have large metabolic heat values their volume will limit their growth so the hyperbolic law should be updated to include other parameters such as the size of the tumor and stage of cancer for example.

From the above stated it can be concluded that one of the important contributions of digital infrared imaging to breast cancer is the possibility of estimating in a non-invasive way the metabolic heat production of a tumor, however further research should be done to determine the influence of external factors such as the accuracy of the tumor depth due to the mammography procedure which deforms the breast, also a more accurate relationship for the metabolic heat generated by breast tumors and their malignancy should be obtained.

## Acknowledgments

The author would like to acknowledge Jesus San Miguel and Julio Castelo from Hospital Central "Ignacio Morones Prieto" who contributed to the breast infrared imaging protocol and Edgar Briones which helped with finite element simulations, also Enrique Martín del Campo and Daniel Marroquin contributed with fruitful discussions regarding digital infrared imaging and breast cancer detection. This work was supported by Fondos Mixtos de San Luis Potosi (FOMIX-SLP) through grant FMSLP-2008-C01-87127.